# Influence of chemical kinetics on detonation initiating by temperature gradients in methane/air

Cheng Wang[1], Chengeng Qian[1], JianNan Liu[1,2] and Mikhail A. Liberman[3]*

[1] State Key Laboratory of Explosion Science and Technology, Beijing Institute of Technology, Beijing, 100081, China
[2] College of Mining Engineering, Taiyuan University of Technology, Taiyuan, 030024, China
[3] Nordic Institute for Theoretical Physics (NORDITA)
Stockholm University, Roslagstullsbacken 23, 106 91 Stockholm, Sweden

**Abstract**

Different simplified and detailed chemical models and their impact on simulations of combustion regimes initiating by the initial temperature gradient in methane/air mixtures are studied. The limits of the regimes of reaction wave propagation depend upon the spontaneous wave speed and the characteristic velocities of the problem. The present study mainly focus to identify conditions required for the development a detonation and to compare the difference between simplified chemical models and detailed chemistry. It is shown that a widely used simplified chemical schemes, such as one-step, two-step and other simplified models, do not reproduce correctly the ignition process in methane/air mixtures. The ignition delay times calculated using simplified models are in orders of magnitude shorter than the ignition delay times calculated using detailed chemical models and measured experimentally. This results in considerably different times when the exothermic reaction affects significantly the ignition, evolution, and coupling of the spontaneous reaction wave and pressure waves. We show that the temperature gradient capable to trigger detonation calculated using detailed chemical models is much shallower (the size of the hot spot is much larger) than that, predicted by simulations with simplified chemical models. These findings suggest that the scenario leading to the deflagration to detonation transition (DDT) may depend greatly on the chemical model used in simulations and that the Zel'dovich gradient mechanism is not necessary a universal mechanism triggering DDT. The obtained results indicate that the conclusions derived from the simulations of DDT with simplified chemical models should be viewed with great caution.

*corresponding author: mliber@nordita.org  (Mikhail (Misha) Liberman)

# 1. Introduction

Explosions of natural gas-air mixtures frequently occur in coal mines and natural gas pipelines and many other industrial processes. Understanding the causes and mechanisms of such explosions is essential for improving safety measures and minimizing devastating hazards. In the worst case explosions may be accompanied by the transition to detonation resulting in a considerable pressure rise and serious damage. Flame acceleration and possible mechanisms of the deflagration-to-detonation transition (DDT) leading to explosions have been extensively studied experimentally, theoretically and numerically [1-3].

If ignited in a confined area the flame accelerates and may undergo deflagration-to-detonation transition (DDT), which can present significant safety hazards. The fundamental mechanisms and processes by which a local small energy release in the reactive mixture can lead to ignition of different chemical reaction modes is one of the most important and fundamental problems in combustion physics. One needs to know how combustion starts and how the transient energy deposition influences the regime of the reaction wave which propagates out from a finite volume of reactive gas where a transient thermal energy was deposited (the hot spot) [4]. This is important for improving safety measures and for understanding ignition risk assessments of processes where hydrocarbons are oxidized at different initial conditions of concentration, temperature and pressure.

A detonation can be initiated directly e.g. by strong shock waves via a localized explosion where a large amount of energy was released. An interesting possibility of the detonation initiation in hydrogen-air caused by focusing of shock waves reflected inside a wedge was recently studied by Smirnov et al [5]. However, in practical cases explosions almost universally start with the ignition of small flame in a relatively small area of combustible mixture where the ignition was initiated either by an electrical spark or another weak ignition source.

A flame ignited near the closed end of a tubes, accelerates and produces pressure waves which steepen into shocks in the flow ahead of the flame front. Various scenarios including shock waves, reflection of shock waves, viscous heating in the boundary layer can lead to the formation of hot spots with an inhomogeneous temperature or reactivity.

Methane-air explosions, and the transition to detonation in such explosions is an intricate problem. Due to the complexity of chemical kinetics for methane/air, until recently a common approach to study DDT has been simulations with a one-step Arrhenius chemistry model. A one-step model of chemical reaction is widely used for numerical simulations of flame dynamics for different geometry of channels (wide or thin) and for obstacle-laden channels in order to understand the mechanism of transition to detonation. The conclusion derived from these simulations was that that the accelerating flame lead to the formation of hot spots, which can produce a detonation through the Zel'dovich gradient mechanism (e.g. [2, 3, 6, 7, 8]). To justify this approach Kessler et al. [8] argued that: "for many practical situations, an extensive description of the details of the chemical pathways is unnecessary. Instead, it is more important to have an accurate model of the fluid dynamics coupled to a model for the chemical-energy release that puts the released energy in the ''right'' place in the flow at the ''right'' time". However, since the ignition times for a one-step model is by orders of magnitude shorter than the experimentally measured and calculated from detailed chemical models ignition times, the fluid dynamics model inevitably puts the released energy in the wrong place at the wrong time, no matter how accurate is the fluid dynamics model. It should also be noted that the flame velocity-pressure dependence given by a one-step model also does not agree with the experimentally measured velocity-pressure dependence. Therefore, results of the simulations of DDT with simplified chemical models should be considered with great caution.

For the first time possible regimes of propagating chemical reaction wave ignited by the initial temperature gradient were studied by Zel'dovich et al. [9] using a one-step Arrhenius model. The Zel'dovich's concept [10] of the spontaneous reaction wave propagating through a reactive mixture along a spatial gradient of reactivity is of great fundamental and practical importance. It opens an avenue to study the reaction ignition and different regimes of the reaction wave propagation initiated by the initial non-uniformity in temperature or reactivity caused by the local energy release.

In a region with nonuniform distribution of temperature the reaction begins at the point of minimum ignition delay time (induction time) $\tau_{ind}(T(x))$ and correspondingly the maximum temperature, and then it spreads along the temperature gradient by spontaneous autoignition at neighboring locations where $\tau_{ind}$ is longer. In the case of a one-step chemical model the induction time is defined by the time-scale of the maximum reaction rate. For the realistic case of a chain branching chemistry this is the time scale of the stage when endothermic chain initiation completed and branching reactions begin. In the case of a one dimension gradient of temperature the spontaneous autoignition wave propagates relative to the unburned mixture in the direction of temperature gradient with the velocity, which is inversely proportional to the gradient of the induction time:

$$U_{sp} = \left| (d\tau_{ind}/dx) \right|^{-1} = \left| (\partial \tau_{ind}/\partial T)^{-1} (\partial T/\partial x)^{-1} \right|. \qquad (1)$$

Since there is no causal link between successive autoignitions, there is no restriction on the value of $U_{sp}$, which depends only on the steepness of temperature gradient. It is obvious, that a very steep gradient (hot wall) ignites a flame, while a zero gradient corresponds to uniform explosion, which occurs in the induction time. For a finite value of the temperature gradient Zel'dovich and co-workers have shown that a sufficiently shallow initial temperature gradient can ignite a detonation regime of combustion [9]. The velocity of spontaneous wave initiated by the temperature gradient decreases while the autoignition wave propagates along the gradient, and reaches the minimum value at the point close to the cross-over temperature where it can be caught-up and coupled with the pressure wave, which was generated behind the high-speed spontaneous wave front due to the chemical energy release. As a result, the pressure peak is formed at the reaction front, which grows at the expense of energy released in the reaction. After the intersection of the spontaneous wave front and the pressure wave, the spontaneous wave transforms into combustion wave and the pressure wave steepens into the shock wave. After the pressure peak becomes large enough, it steepens into a shock wave, forming an overdriven detonation wave.

Obviously, a simplified one-step and even more advanced two-step or four-step models, which to some extent mimic the chain-branching kinetics, do not describe properly systems governed by a large set of chain-branching reactions. The quantitative and in some cases qualitative difference between simplified models and detailed chemical models remained unclear. Liberman et al. [11, 12] employed detailed chemical kinetic schemes to study the combustion regimes in stoichiometric hydrogen-oxygen and hydrogen-air mixtures ignited by the initial temperature gradient. It was shown that the evolution of a spontaneous wave calculated using the detailed kinetic model is qualitatively different compared with the predictions obtained from calculations with a one-step model. First, the induction times predicted by a one-step Arrhenius model are in orders of magnitude smaller than the induction times predicted by a detailed chemical model. Another difference is that for the one-step model the reaction is exothermic for all temperatures, while chain branching reactions start with endothermic induction stage representing chain initiation and branching. Therefore, for a detailed model the hydrodynamics is effectively "switched-off" during induction stage. As a consequence, combustion regimes initiated by the temperature gradient require much shallower gradients compared with those predicted by a one-step model. This means that the size of a hot spot with a temperature gradient capable of producing detonation obtained in simulations with a detailed chemical model can be by orders of magnitude greater than that obtained from simulations with a one-step model. The size of a hot spot with a temperature gradient capable of producing detonation decreases with the increase of initial pressure [12], and may become rather small at very high pressure for hydrogen-air [13].

From the results of simulations with a one-step model, the Zel'dovich gradient mechanism is often considered as a universal mechanism explaining the transition from deflagration to detonation [2, 3, 6, 7, 8]. This trend was considered as the mainstream in DDT studies until it was shown experimentally by Kuznetsov et al. [14] that for a stoichiometric hydrogen/oxygen and ethylene-air mixtures the temperature in the vicinity of the flame prior to DDT remains too low (does not exceed 550K) for spontaneous ignition. Experimental studies and numerical simulations of DDT with the detailed chemical model for hydrogen/oxygen [15-18] have shown that the gradient mechanism cannot explain DDT. A new mechanism of DDT consisting in the mutual amplification of a weak shock formed very close ahead of the flame front and coupled with the flame reaction zone was proposed by Liberman et al. [14]. This mechanism of DDT is basically very similar to the SWACER mechanism (shock wave amplification by coherent energy release) proposed by Lee and Moen [19], and the similarity of the SWACER mechanism and coherent amplification of the shock wave and the flame reaction is the probable reason that the results of simulation with a one-step chemical model were often interpreted as the spontaneous wave formation following by the onset of detonation via the gradient mechanism.

Since the use of detailed chemical mechanisms can severely limit the calculations, the reduced chemical models with a minimum number of reactions and species, suitable for describing transient combustion processes, are of great interest, especially for methane-air combustion, where complete reaction mechanism can consist of many hundred reactions and species. In the present work we describe the first steps toward achieving the reliable kinetic model, suitable for multidimensional simulations of different stages of the methane-air combustion, including the flame acceleration and detonation initiation. The reliability and fidelity of different reduced chemical models for methane-air mixtures are tested. We then use the detailed chemical models with a minimum number of reactions and species but detailed enough to describe correctly the most important for multidimensional simulations characteristics of the flame: the laminar flame velocity and structure, the ignition delay times for a wide range of initial pressures and temperatures, and the dependence of the flame velocity on pressures. It is shown that there is considerable difference in the detonation initiation by temperature gradient for a detailed chemical reaction model compared to simplified models. The size of a hot spot L, which can be viewed as the inverse steepness or length of the temperature gradient, $L = T / (dT / dx)$, capable of producing detonation obtained with the detailed chemical model might be much larger than that obtained from calculations with a one-step or other simplified models.

**1. Chemical kinetics modeling**

In the present paper we compare different chemical models for methane-air combustion. The reliability of a chemical model to be used for simulations requires that the model predicts correctly the laminar flame velocity and structure and the ignition delay times at different pressures. The flame velocity, induction times, and their pressure dependence, calculated for different chemical models, were compared with calculations using the standard GRI Mech 3.0 and experimental data. These requirements stem from the need of ensuring the controllable capability of characterizing ignition for transient combustion processes evolving over wide ranges of temperature and pressure in the processes of the flame acceleration and DDT. The capability of the model to predict correctly laminar flame speeds and structure at normal conditions is not sufficient, since the flame propagating in confined area (in a channel) and the spontaneous reaction waves are accompanied by pressure and shock waves. This means that the model must predict correctly the dependence of a laminar flame speed on pressure. Finally, the computational cost to introduce the chemical schemes into DNS or LES solvers must be reasonable for the feasibility of multidimensional CFD simulations.

*2.1. Simplified reaction mechanisms: Single-step, two-step and four-step mechanisms*

A single-step chemistry approach suggests that the complex set of reactions can be modeled by a one-step Arrhenius reaction and has been widely used for theoretical studies and simulations [1, 2, 3]. We will test the one-step Arrhenius model, which was used by Kessler et al. [8] for 2D simulations of the methane–air flame acceleration and DDT in channel with obstacles at initial pressure $P = 1 atm$ and temperature $T = 298 K$

$$W = A\rho Y \exp(-E_a / RT). \qquad (2)$$

The parameters used in [8] to calibrate the model are: the pre-exponential factor, $A = 1.64 \cdot 10^{13} cm^3 / g s$; the activation energy of the reaction $E_a = 67.55 RT_0$, density of the gas, $\rho = 1.1 \cdot 10^{-3} g / cm^3$, the ratio of specific heats, $\gamma = C_P / C_V = 1.197$, R is the universal gas constant, $T_0 = 298 K$.

Following the idea proposed for the first time by Zeldovich [20], different versions of two-step models have been developed to mimic chain branching character of the reaction [21]. Two-step models contain only reactions of hydrocarbons oxidation, which are the most important for the heat release. Slow reactions at T< 800 K with a small heat release are not considered. As a representative example, we consider the widely used versions of two-step models: 2S-CH$_4$ mechanism developed by Westbrook et al. [22] and 2S-CH$_4$-BFER developed by Franzelli et al. [23]. The 2S-CH$_4$-BFER model is an improved version of the classical two-step mechanism [22], and was proved to be valid for lean and stoichiometric mixtures, providing an accurate description of the chemical flame structure and velocity. The 2S-CH4-BFER mechanism [23] uses 6 species ($CH_4$, $O_2$, $CO_2$, CO, $H_2O$, $N_2$) for 2 global reactions and correctly predicts the chemical flame structure and velocity of a premixed laminar methane/air flame for a wide range of equivalence ratio and pressures ($1 \text{ atm} \leq P_0 \leq 10 \text{ atm}$) and for fresh mixture temperatures ($300 \text{ K} \leq T_0 \leq 800 \text{ K}$). A further improvement of the reaction mechanism into several steps, the 4-step reaction scheme developed by Jones and Lindstedt [24]. The scheme consists of four global reactions describing chain branching, chain breaking and CO oxidation: I) $CH_4 + 2H + H_2O \Leftrightarrow CO + 4H_2$, II) $CO + H_2O \Leftrightarrow CO_2 + H_2$, III) $H + H + M \Leftrightarrow H_2 + M$, IV) $3H_2 + O_2 \Leftrightarrow 2H + 2H_2O$. The 4-step reaction scheme makes the predicted product temperature and composition more accurate, though it costs more computation time.

*2.2 Comparison of simplified and detailed chemical schemes*

It is known that to be able to predict correctly the ignition delay times the reduced mechanisms must consist of at least ten reactions [25]. We compare predictions of the simplified models with a reduced detailed reaction model DRM-19 developed by Kazakov and Frenklach [26], which consists of 19 species and 84 reactions, and with the 35 reactions model, developed by Smooke [27]. These models are selected due to lesser number of species and ensure reasonably faster computations. The reaction mechanism DRM-19 has been chosen as a main one for simulating as it was extensively validated by many researchers for combustion characteristics related to ignition delay times and laminar flame velocities over a wide range of pressures, temperatures, and equivalence ratios [28, 29, 30]. The results for all chemical models were compared with the standard detailed reference mechanism GRI 3.0 Mech [31] and experimental data. The capability of different reaction schemes to reproduce the laminar flame structure and speeds and the ignition delay times were examined using high resolution simulations. The resolution and the grid independence (convergence) tests, in particularly for shock wave, were thoroughly performed and presented in Appendix A.

The ignition delay time can be defined as the time during which the maximum rate of temperature rise ($\text{Max}\{dT/dt\}$) is achieved, which is close to the time of the exothermic reactions activation. The ignition delay times were calculated for different chemical schemes using the standard constant volume adiabatic model. While all the simplified models allow to reproduce characteristics of the laminar flame (the laminar flame speed $U_L$ and the adiabatic flame temperature $T_b$) with satisfactory good accuracy, the induction times predicted by a one-step and other simplified models are more or less close to each other but differ significantly from the values calculated using the detailed chemical models, GRI 3.0 Mech, and the experimentally measured ignition delay times.

Figure 1 shows the induction times versus temperature at the initial pressure 1 atm computed using simplified and detailed chemical reaction models, as well as the induction times calculated using GRI 3.0 Mech and some recent experimental results [32, 33]. It is seen that the induction times predicted by detailed chemical models are in a good agreement with GRI 3.0 Mech and with the experimental results [32, 33]. The induction times predicted by simplified chemical models is up to three orders of magnitude shorter than that predicted by detailed chemical models. The curve $\tau_{ind}(T)$ calculated for DRM-19 and GRI 3.0 Mech for temperatures in the range 800-1000K is shown by dashed line because of uncertainty of the "cool flames" contribution. At the same time, this low temperature part of the induction time is not important for the present study.

Since a one-step and simplified models predict much shorter induction times than the induction times predicted by detailed chemical models, they effectively much earlier "switched-on" the gasdynamics during induction stage. Therefore, the spontaneous wave velocity produced by the same temperature gradient is considerably different for the detailed chemical models compared to that computed using the one-step or simplified models. Correspondingly, the temperature gradient capable to ignite a detonation is expected to be much shallower for the detailed chemical models than that computed with simplified models.

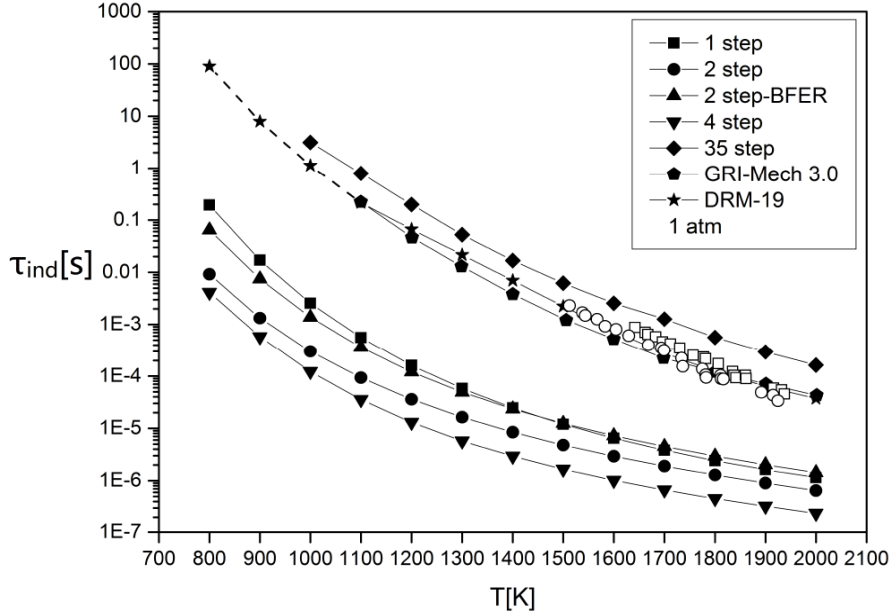

**Figure 1:** Induction times for stoichiometric methane-air mixture at pressure $P = 1$ atm calculated at different temperatures using simplified and detailed chemical kinetic schemes. Empty circles and squares are experimental data [32] and [33], correspondingly.

This trend of a very large difference between the induction times calculated for simplified models and the actual ignition times measured experimentally and calculated using detailed chemical models remains almost the same at elevated pressures. The difference in induction times for different simplified models is not so large. It will be shown below that this difference leads to some noticeable but not very strong difference in the steepness of the temperature gradients for the initiation of detonation. Another difference between the induction times calculated for simplified models and for the detailed model is the different slope of the curves $\tau_{ind}(T)$. This difference may affect considerably the value of spontaneous wave, which is according to Eq. (1) proportional to $(d\tau_{ind} / dT)^{-1}$. Figures 2 and 3 show the induction times calculated for the initial pressure 5 atm and 10atm for simplified and detailed models DRM-19 and GRI3.0 with empty symbols showing the experimental results [32, 33].

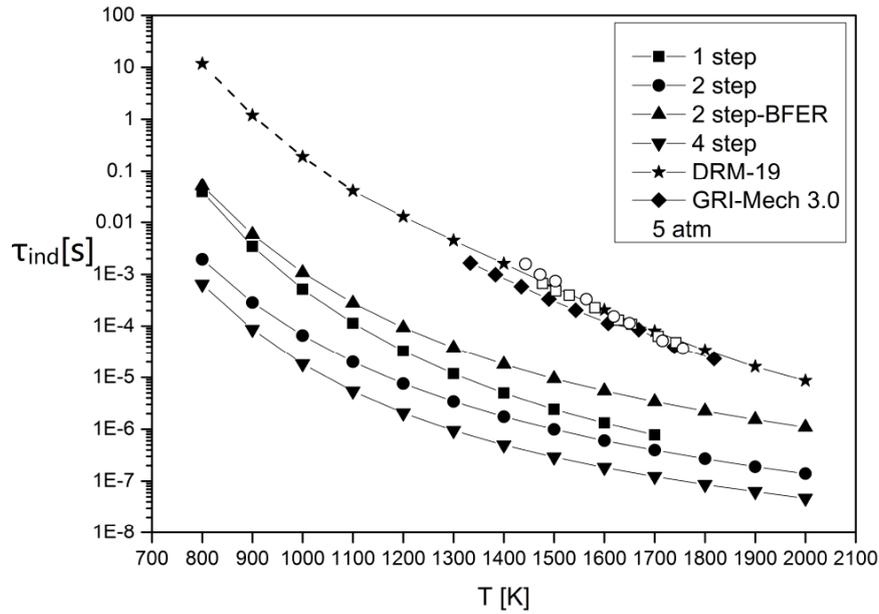

**Figure 2**: Comparison of induction times for methane-air at initial pressure 5atm, calculated for: 1-step, 2-step, 4-step models, the detailed model DRM-19 and GRI.3.0. Empty circles and squares are experimental results [32] and [33].

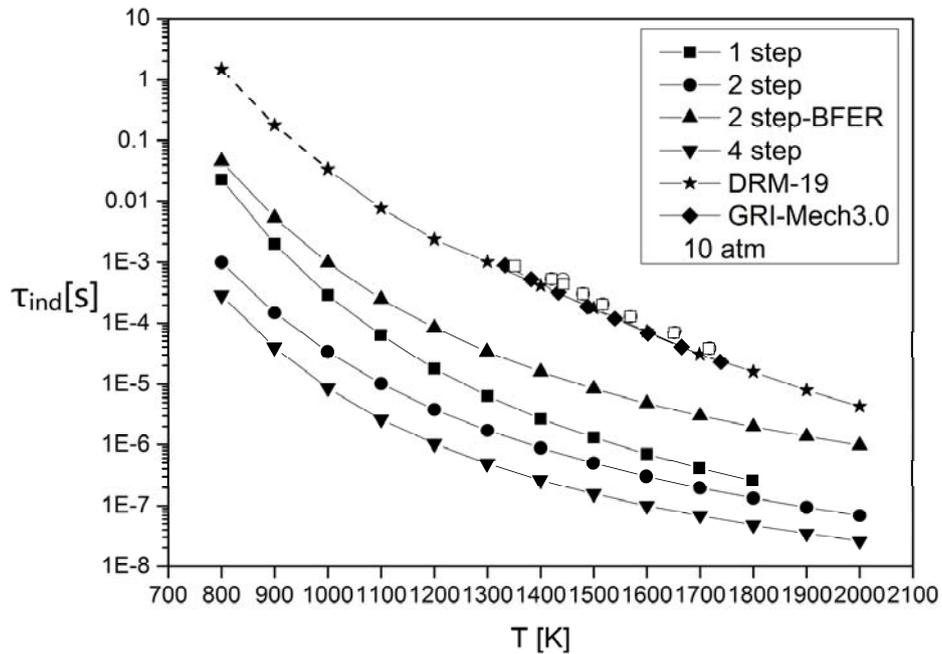

**Figure 3**: Comparison of induction times for methane-air at initial pressure 10 atm, calculated using a 1-step, 2-step, 4-step models, the detailed model DRM-19 and GRI.3.0. Empty circles and squares are experimental data [32] and [33].

From the physical point of view, the difference between the induction times calculated using simplified models from the induction times, measured experimentally and calculated using detailed chemical models, means that the calibration of simplified models corresponds to the activation energy of 6-8 times smaller than the effective activation energy.

For all the chemical models considered in the present paper the 1D simulations of a plane flame performed using the DNS solver for the fully compressible multispecies Navier-Stokes equations predict satisfactory good agreement for the flame velocity at normal conditions, $P_0 = 1$ atm.

Another shortcoming of simplified models, especially the one-step model is a strong discrepancy of the laminar flame velocity-pressure dependence predicted by a one-step model compared to the velocity-pressure dependence calculated using the DRM19 and measured in experiments.

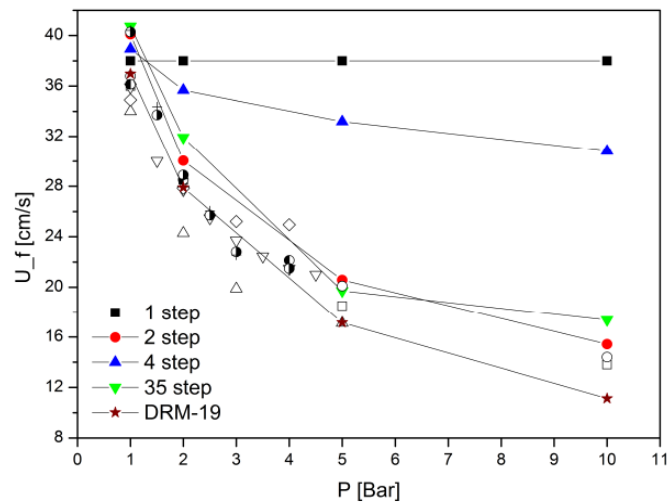

**Figure 4:** Laminar flame velocities vs pressure calculated using simplified and detailed chemical models. Symbols on the figure are the experimental results from: ▽ --[28]; □-- [34]; ○-- [35]; △ --[36]; ◇ -- [37]; ☉-- [38]; ☽-- [39]; + --[40].

Figure 4 shows the laminar flame velocity-pressure dependence calculated for simplified and detailed chemical models. The experimental measurements [28, 34, 35, 36, 37, 38, 39, 40] in the figure 4 are shown by symbols. With the increasing pressure the flame velocity decreases and becomes about 3 times smaller at 10 atm in agreement with predictions of DRM19 and experimental measurements. On the contrary, a one-step model predicts a constant, pressure-independent velocity of the laminar flame.

**3. Influence of chemical models on detonation initiation by the temperature gradient**

In what follows we will use the conventional term hot spot as a location within a reactive mixture where temperature is higher than in surrounding mixture. This means that the hot spot size is viewed as the scale (inverse steepness), $L = T/(dT/dx)$) of the temperature gradient.

*3.1. Problem setup*

We consider uniform initial conditions apart from a linear temperature gradient. The model of the linear temperature gradient is convenient for analysis and it has been widely used previously by many authors [9, 10, 11, 12, 13, 41-46]. The emphasis will be on quantitative and qualitative comparison and contrasting the results obtained from the modeling obtained for simplified chemical models and for the detailed DRM19 chemical model.

The initial conditions at $t = 0$, prior to ignition, are constant pressure and zero velocity of the unburned mixture. At the left boundary, at $x = 0$, the conditions are for a solid reflecting wall, where $u(0,t) = 0$ and the initial temperature, $T(x=0) = T*$ exceeds the ignition threshold value. Thus, the initial conditions are quiescent and uniform, except for a linear gradient in temperature (and hence density):

$$T(x,0) = T* - (T* - T_0)(x/L), \text{ at } 0 \leq x \leq L. \tag{3}$$

$$P(x,0) = P_0, \; u(x,0) = 0. \tag{4}$$

The initial temperature gradient is characterized by the temperature $T(0,0) = T*$ at the left end, by the uniform temperature of the mixture outside the gradient, $T(x > L, 0) = T_0$ and by the gradient steepness, $(T* - T_0)/L$. The "gradient length", $L = (T* - T_0)/(dT/dx)$, which characterizes the gradient steepness can be viewed as the size of the hot spot, where the initial temperature gradient is formed.

The 1D direct numerical simulations were performed using the DNS solver, which used the fifth order weighted essentially non-oscillatory (WENO) finite difference schemes [47] to resolve the convection terms of the governing equations and the sixth order standard central difference scheme to discretize the non-linear diffusion terms. The advantage of the WENO finite difference method is the capability to achieve arbitrarily high order accuracy in smooth regions while capturing sharp discontinuity. The time integration is third order strong stability preserving Runge-Kutta method [48]. The DNS solver was used to solve the set of the one-dimensional time-dependent, fully compressible reactive Navier-Stokes equations and chemical kinetics.

$$\frac{\partial \rho}{\partial t} + \frac{\partial (\rho u)}{\partial x} = 0, \tag{5}$$

$$\frac{\partial \rho u}{\partial t} + \frac{\partial (\rho u^2 + P)}{\partial x} = \frac{\partial \tau_{xx}}{\partial x} \tag{6}$$

$$\frac{\partial \rho E}{\partial t} + \frac{\partial (\rho E + P)u}{\partial x} = \frac{\partial}{\partial x}(\tau_{xx} u) + \frac{\partial}{\partial x}\left(\kappa(T)\frac{\partial T}{\partial x} + \rho \sum_{i=1}^{N_S} h_i Y_i V_{i,x}\right) \tag{7}$$

$$\frac{\partial \rho Y_i}{\partial t} + \frac{\partial \rho Y_i (u + V_{i,x})}{\partial x} = \omega_i. \tag{8}$$

The equation of state is

$$P = \rho T \sum_{i=1}^{N_S} Y_i R_i = \rho T R_B \sum_{i=1}^{N_S} Y_i / W_i. \tag{9}$$

Here we use the standard notations: $\rho, u, P, E, R_B$ and $T$, are density, flow velocity, pressure, specific total energy, the universal gas constant and temperature. $Y_i, R_i, V_{i,x}, \omega_i$ and $W_i$ are the mass fraction, specific gas constant, diffusion velocity, reaction rate and the molar weight of i-species. $h_i$ is specific enthalpy of species, $R_B$ - is the universal gas constant.

The specific heats and enthalpies of each species are taken from the JANAF tables (Joint Army Navy NASA Air Force Thermochemical Tables) and interpolated by the fifth-order polynomials [49]. In the case of a single-step model the ideal gas equation of state was used. The viscosity coefficients of the mixture were calculated by employing the empirical equation

$$\mu = \frac{1}{2}\left[\sum_{i=1}^{N_s} X_i \mu_i + \left(\sum_{i=1}^{N_s} \frac{X_i}{\mu_i}\right)^{-1}\right], \tag{10}$$

where $X_i$ is the molar fraction and $\mu_i = \frac{5}{16}\frac{\sqrt{\pi m_i kT}}{\pi \sigma_i^2 \tilde{\Omega}_i^{(2,2)}}$ is the coefficient of viscosity of specie i, $m_i$, $\sigma_i$ and $\tilde{\Omega}^{(2,2)}$ are the molecule mass, molecule diameter and the collision integral, which is from the Lennard-Jones potential [50]. Coefficients of the heat conduction of specie i were calculated from Eucken formula

$$\kappa_i = \left(c_{p,i} + \frac{5}{4}R_i\right)\mu_i, \tag{11}$$

which is similar to $\kappa_i = \mu_i c_{pi}/\Pr$ with the Prandtl number, $\Pr = 0.75$. The averaged coefficient of the heat conduction of specie i is

$$\kappa = \frac{1}{2}\left(\sum_{i=1}^{N_s} X_i \kappa_i + \left(\sum_{i=1}^{N_s} \frac{X_i}{\kappa_i}\right)^{-1}\right) \tag{12}$$

The diffusion coefficient of i-th species is

$$D_i = \frac{1-Y_i}{\sum_{j \neq i} X_i/D_{ij}}, \tag{13}$$

where $D_{ij}$ is the binary diffusion coefficients [15, 16, 17].

$$D_{ij} = \frac{3}{8}\frac{\sqrt{2\pi kT\, \hat{m}_i \hat{m}_j/(\hat{m}_i + \hat{m}_j)}}{\pi \cdot \rho \cdot \sigma_{ij}^2 \tilde{\Omega}^{(1,1)}} \tag{14}$$

The concentrations of the mixture components $Y_i$ are defined by the solution of system of chemical kinetics.

*3.2. Numerical simulations: Simplified chemical models*

In this section we consider the critical size of the hot spot capable to initiate detonation in the methane-air stoichiometric mixture by the temperature gradient with initial conditions given by Eqs. (3) and (4) for the simplified: 1-step, 2-step and 4-step chemical models.

It was shown by Liberman et al. [11, 12] for the highly reactive hydrogen-oxygen that the longer delay associated with the induction stage of reaction for detailed chemical model, the faster initial speed and weak chemical-acoustic adjustment during the early phase of evolution result in considerably different spontaneous wave speeds.

In general, classification of possible modes of the propagating combustion wave inspired by the spontaneous wave initiated by the temperature gradient is similar to that described by Liberman et al. [12] for the highly reactive stoichiometric hydrogen-oxygen. The pressure waves generated during the exothermic stage of reaction can couple and evolve into a self-sustained detonation wave, or produce a flame and a decoupled shock, depending on the gradient steepness and on the value of temperature $T_0$ outside the hot pocket. The outcome depends on the gradient steepness and the relationship between the speed of the spontaneous wave at the point where its velocity reaches a minimum, $\min\{U_{sp}\}$ and the characteristic velocities of the problem: the laminar flame speed $U_f$, the speeds of sound at the points $T^*$ and $T_0$: $a_s(T^*)$ and $a_s(T_0)$, speeds at the Newman point $a_N$, at the Chapman-Jouguet point, $a_{CJ}$ and the velocity of Chapman-Jouguet detonation $U_{CJ}$. Because of the limited space, below we will focus on the most interesting aspect - the conditions under which the temperature gradient can initiate a steady detonation.

The velocity of the spontaneous wave initiated by the initial temperature gradient decreases while the wave propagates along the gradient, and reaches its minimum value, $\min\{U_{sp}\}$, at the point close to the crossover temperature. Therefore, the necessary condition for initiating detonation by the spontaneous reaction wave is that the spontaneous wave initiated by the initial temperature gradient can be caught up and coupled with the pressure wave which was generated behind the high-speed spontaneous wave front. This means that the necessary condition for initiating detonation by the temperature gradient can be written as

$$U_{sp}(T_{cr}) = \left(\frac{\partial \tau}{\partial T}(T_{cr})\right)^{-1}\left(\frac{\partial T}{\partial x}(T_{cr})\right)^{-1} = \left(\frac{\partial \tau}{\partial T}(T_{cr})\right)^{-1}\frac{L}{T^* - T_0} \geq a_s(T_{cr}) \qquad (15)$$

where $T_{cr}$ is temperature at the point x corresponding to $\min\{U_{sp}\}$. Using this condition we can estimate the maximum steepness of the temperature gradient, or the critical gradient length (the minimum size of the hot pocket, $L_{cr}$) for successful detonation initiation.

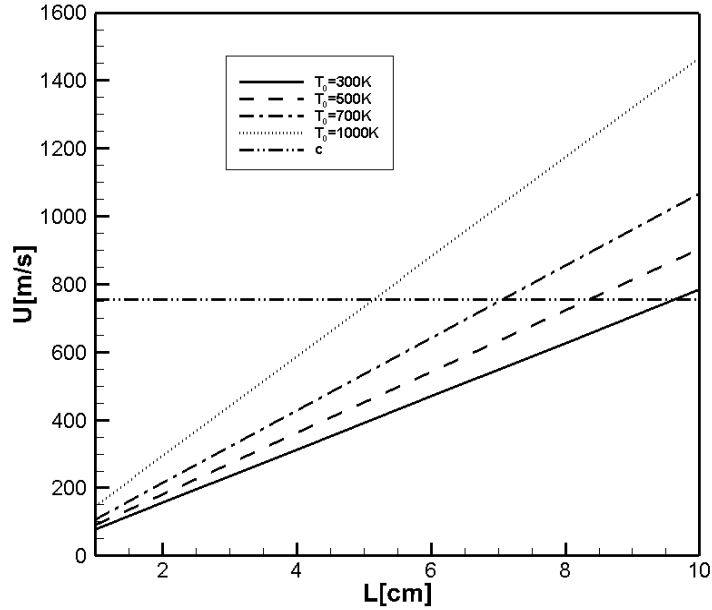

**Figure 5:** The points of intersection of lines $U_{sp}(T_{cr})$ with the sound speed line correspond to the steepest gradients $L = L_{cr}$. The "crossover" temperature, $T_{cr} = 1500K$, $T_0 = 300; 500 : 700; 1000$ K; the horizontal dot-dashed line is the sound speed at $T = 1500K$.

Figure 5 shows $\min\{U_{sp}\}$ for a one-step chemical model, $P_0 = 1\text{bar}$, $T^* = 1800K$. The values of L, for which the lines $U_{sp}(L)$ from (Eq. 15) intersect the line $a_s(T_{cr})$, correspond to the minimum size of the hot packet $L_{cr}$ at which the synchronization and mutual amplification between the travelling shock (pressure) wave and reaction front associated with the self-ignition mechanism at lower temperature, amplifies the shock wave and the spontaneous transition to detonation occurs depending on the initial temperature $T_0$ outside the hot pocket. In the case of a one-step model reaction is exothermal for any temperature and there is not "crossover" temperature separating induction and exothermal stages. In this case, $T = 1500K$ was taken as the temperature, corresponding location of the spontaneous wave, where energy previously released in the reaction resulted in the first noticeable pressure peak.

According to the diagram in Fig.5 a steady detonation can be developed by the temperature gradient ($P_0 = 1\text{bar}$, $T^* = 1800K$, $T_0 = 300K$) if it steepness corresponds to $L \geq L_c \approx 9\text{cm}$.

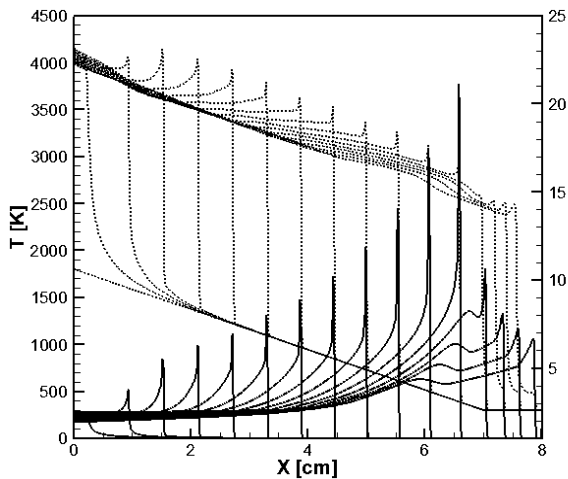 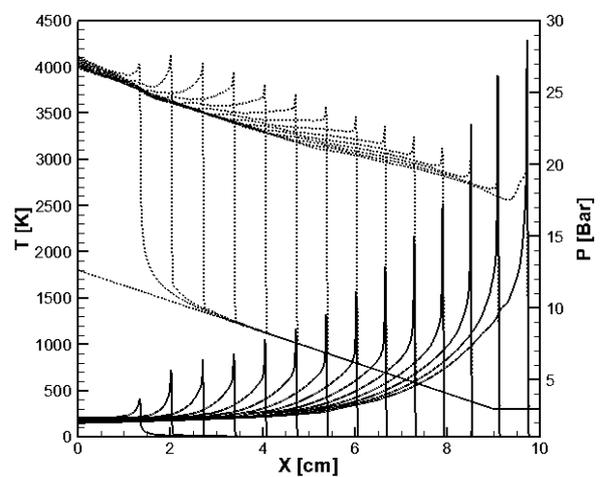

**Fig. 6(a)**                                   **Fig. 6(b)**

**Figure 6(a, b):** Evolution of the temperature (dashed lines) and pressure (solid lines) profiles during the formation of the detonation at $P_0 = 1\text{atm}$: Fig.6a: $L = 7\text{cm}$, $T_0 = 300K$; Fig.6b: $L = 9\text{cm}$, $T_0 = 300K$.

Figure 6(a) shows evolution of the temperature and pressure profiles during the formation and quenching of detonation calculated for a one-step chemical model Eq.(2), $P_0 = 1\text{bar}$, $T^* = 1800K$, $T_0 = 300K$, for the "steep" temperature gradient, $L = 7cm < L_{cr}$. In this case the reactive zone starts to move slowly away from the leading shock wave. The rarefaction wave propagates into the reaction zone and the separation between the zone of heat release and the leading shock increases. As a result, the intensity of the shock wave becomes weaker and detonation quenches. On the contrary, Figure 6(b) shows development of steady detonation for shallower gradient, $L \geq L_{cr} \approx 9cm$.

The evolution of the reaction wave and pressure peak (shock wave) velocities for the conditions of Fig. 6(a) and Fig. 6(b) is shown in Figs. 7(a, b). The velocity of the reaction wave was calculated from the trajectory of the reaction front and the velocity of the pressure wave was calculated from the trajectory of the maximum pressure of the pressure wave profile.

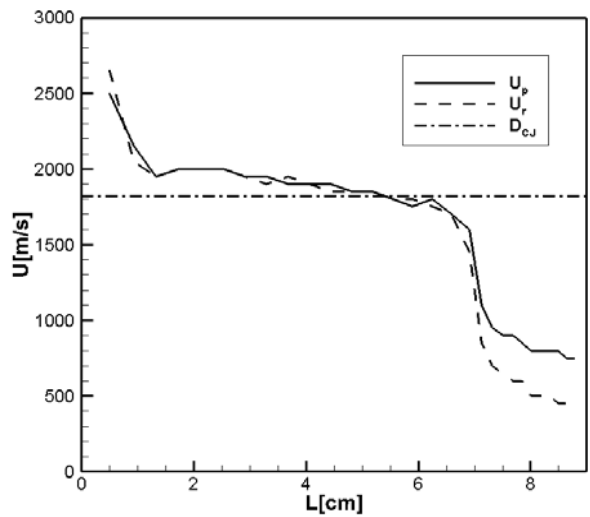 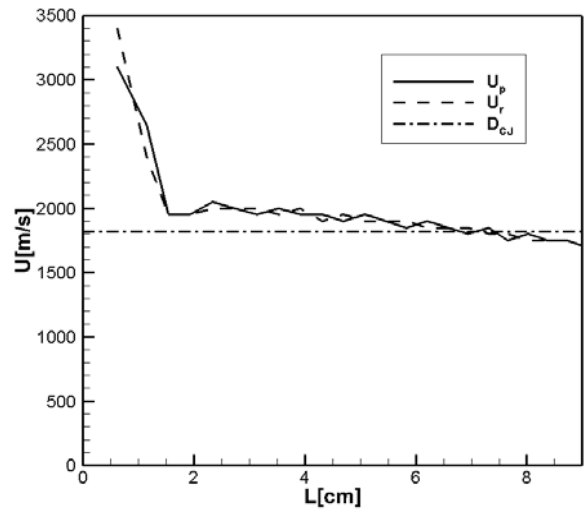

**Fig. 7(a)**                                          **Fig. 7(b)**

**Figure 7(a, b):** Velocities of the spontaneous wave (solid line) and pressure wave (dash-dotted lines) computed for the conditions in Fig.6(a) and Fig.6(b).

If the initial temperature gradient is steeper ($L = 7cm$ in Fig. 6(a)), the reaction front velocity at the point, where the pressure wave overtakes the reaction wave, is not sufficient to sustain synchronous amplification of the pressure pulse in the flow behind the shock wave. As a result, the pressure wave runs ahead of the reaction wave and the velocity of the reaction wave decreases (Figs 6(a), 7(a)). On the contrary, for the shallower gradient, $L \geq L_{cr} \approx 9 cm$, a steady Chapman-Jouguet detonation develops with the velocity $D_{CJ}$ shown in Fig.7(b).

For the first time, phenomenon of spontaneous quenching of the developing detonation has been studied by He and Clavin [41, 42, 43]. This led to definitions of critical sizes of initial hot pockets of fresh mixture for igniting a detonation, observed by He & Clavin [42]. They also pointed that for a given temperature T* the local criterion $U_{sp}(x) \sim a*(x)$ for the spontaneous formation of a CJ detonation defines a critical temperature gradient. It should be noted that criterion Eq. (15) more accurately gives a critical gradient and its dependence on the temperature outside a hot pocket. It is seen that for steeper gradient ($L = 7 cm$) shown in Fig. 6(a) detonation developed from the beginning, follows closely the local CJ detonation in the early stage of propagation but later on it quickly quenches at $x > 6.5 cm$.

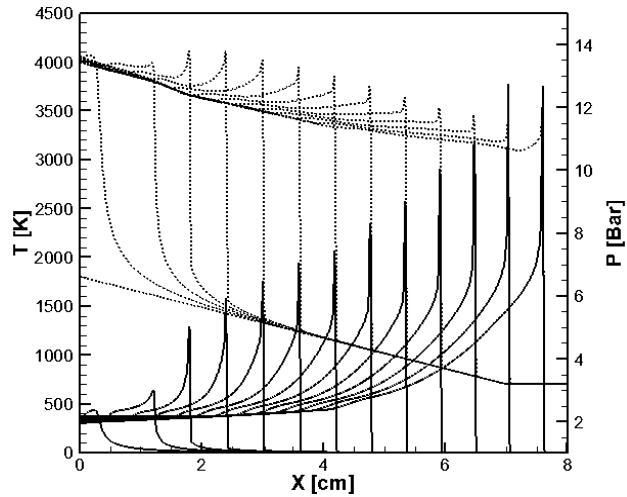

**Figure 8:** Evolution of the temperature (dashed lines) and pressure (solid lines) profiles during the detonation formation: $P_0 = 1 atm$  $L = 7 cm$, $T_0 = 700 K$.

Later on there is the departure from the CJ detonation and then detonation quickly quenches. He and Clavin [41] also emphasized that the same temperature gradient, for which there is a detonation quenching, can ignite a detonation at a higher temperature $T_0$ outside the hot pocket. Indeed, in agreement with diagram in Fig. 5, Figure 8 shows that for the same temperature gradient as in Fig. 6a, but for $T_0 = 700K$ outside the hot pocket, detonation does not quench and develops in a steady CJ detonation.

He and Clavin [42] explained spontaneous quenching of detonation in simple terms using a particular form of the quasi-steady-state approximation. In the classification of reaction waves initiated by temperature gradient by Liberman et al. [12] this corresponds to a quasistationary structure consisting of a shock wave and reaction zone, which may transform into a detonation propagating down the temperature gradient for the condition $a_N < \min\{U_{sp}\} < a_{CJ}$.

In general, the classification of the combustion modes initiated by the initial temperature gradient, as well as the critical size $L_{cr}$ of the hot spot capable to initiate detonation, are quite similar for all the simplified: 1-step, 2-step and 4-step chemical models. Figures 9(a, b) show the temporal evolution of the temperature and the pressure profiles during development of a steady detonation calculated for 2-step and 4-step models for the same conditions: $P_0 = 1bar$, $T^* = 1800K$, $T_0 = 300K$. For 2-step model used in calculations shown in Fig. 9(a), the critical size of the hot spot is $L_{cr} = 19cm$, and in calculations with 4-step chemical model, shown in Fig.9(b), the critical size of the hot spot, characterizing steepness of the temperature gradient is $L_{cr} = 9cm$.

It should be noted that although the induction time for the 4-step model is smaller than for the 2-step model, and in both cases $\tau_{ind}$ is less than $\tau_{ind}$ for the one-step model, the critical size $L_{cr} = 9cm$ for the 4-step model is the same as for the 1-step model, and for the 2-step model $L_{cr} = 19cm$ is somewhat larger than for a than for one-step model.

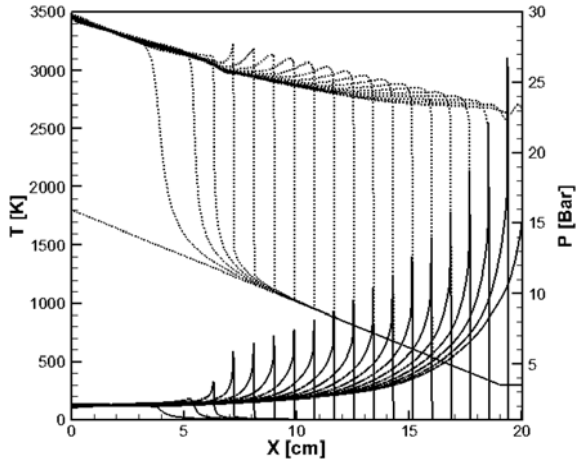 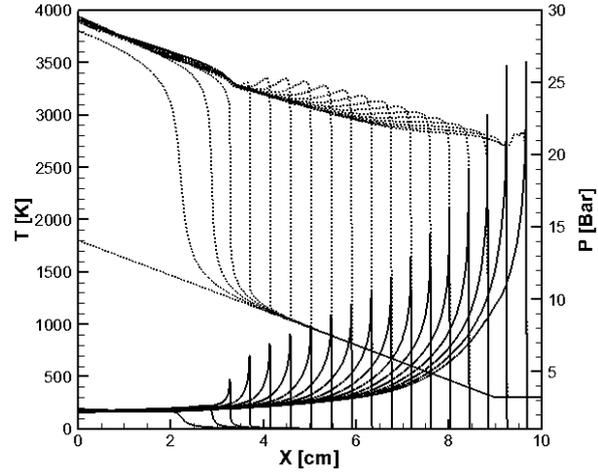

**Fig. 9(a)**          **Fig. 9(b)**

**Figure 9(a, b):** Evolution of the temperature (dashed lines) and pressure (solid lines) profiles during the formation of the detonation for $P_0 = 1\text{bar}$, $T_0 = 300K$. Fig.9a: two-step model, $L_{cr} \approx 19\text{cm}$; Fig. 9b: four-step model, $L_{cr} \approx 9\text{cm}$.

The difference in critical hot spot sizes $L_{cr}$ at which the temperature gradient initiates detonation is due to the fact that $L_{cr}$ depends not only on the induction time, but also on $(\partial \tau_{ind} / \partial T)$ that determines the speed of the spontaneous wave. The effect of "the thermal sensitivity" of the induction time, $\beta = -(T/\tau_{ind})(\partial \tau_{ind} / \partial T)$ will be discussed in more detail in Sec. 5, where the detonation initiation by the temperature gradient is compared for a detailed chemical model with that for the one-step model.

### 3.3. Simplified models. Detonation initiated by temperature gradient at elevated pressures

As the initial pressure increases, induction times decrease for all chemical models as can be seen from Figures 1, 2, 3. Also, as the initial pressure increases, the speed of the spontaneous wave increases rapidly, since $(\partial \tau_{ind} / \partial T)$ decreases. Liberman et al. [12] have pointed that while at low pressures the induction zone is much longer than the chain termination exothermic zone for hydrogen/oxygen, at high pressures, when triple collisions dominate, they become of the same order. The corresponding temperature at which the equilibrium of the induction and termination stages takes place is known as the extended second explosion limit. At higher pressures this limit shifts to higher temperatures. Therefore, at high pressures the scenario of detonation initiation by the temperature gradient for all chemical models should become somewhat more similar to that realized for a one-step model. Since at high pressure the induction time is smaller, the minimal steepness of the gradients necessary for the detonation initiation of increases (the $L_{cr}$ decreases). Evolution of the temperature and pressure profiles during the formation of the detonation calculated for one-step, two-step and four-step models at $P_0 = 5\text{bar}$; $T_0 = 300\text{K}$, is shown in Figures 10, 11(a) and 11(b), correspondingly. Interesting is that already at $P_0 = 5\text{bar}$, the minimal steepness of the gradients necessary for the detonation initiation became the same for all simplified models (the minimal size of the hot pocket, $L_{cr}$=3cm).

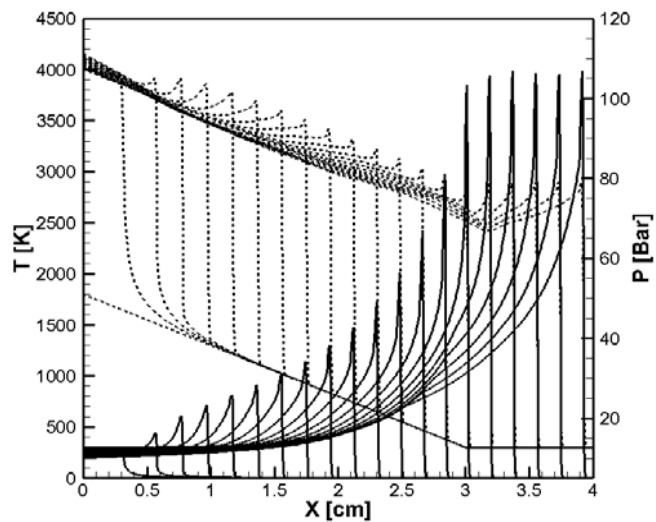

**Figure 10:** Evolution of the temperature (dashed lines) and pressure (solid lines) profiles during the formation of the detonation: one-step model, $P_0 = 5\text{bar}$; $T_0 = 300\text{K}$, $L_{cr} \approx 3\text{cm}$.

Since at high pressure the ranges of "speed" limits separating regions of different modes, which are determined by the sound speeds, $a_0$, $a*$ and $a_{CJ}$ decrease, the ranges for the realization of all combustion modes decrease correspondingly [12]. We consider how the scenario of detonation initiation at high pressures changes for detailed chemical model DRM19 compared to a one-step model in more details in Sec. 3.5.

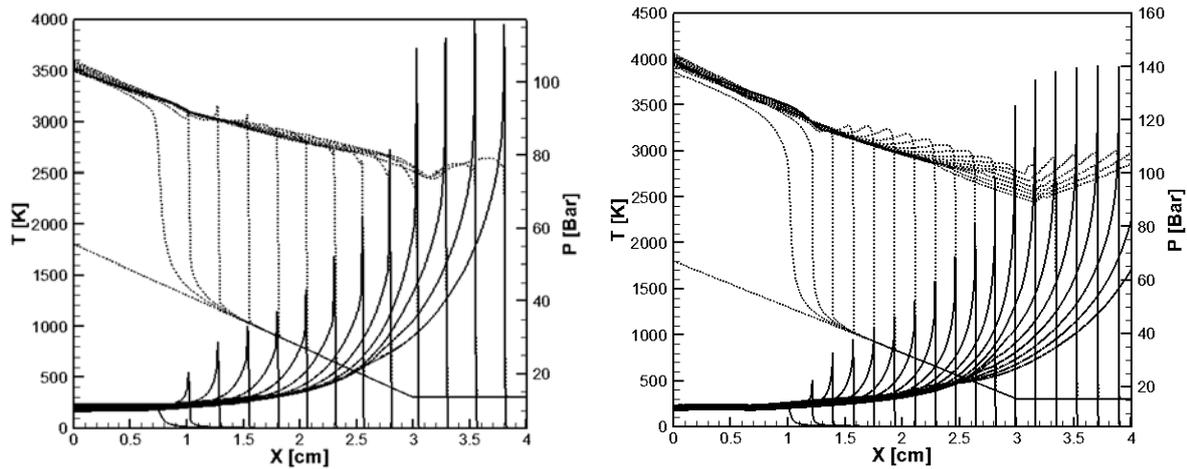

**Figure 11:** Evolution of the temperature (dashed lines) and pressure (solid lines) profiles during the formation of the detonation for $P_0 = 5\,\text{bar}$, $T_0 = 300\,\text{K}$, $T^* = 1800\,\text{K}$. Fig.11a: 2-step model, $L_{cr} \approx 3\,\text{cm}$; Fig.11b: 4-step model, $L_{cr} \approx 3\,\text{cm}$.

### 3.4. Detonation initiated by temperature gradient for detailed chemical model DRM-19

According to the results shown in Figs. 1-3, the ignition delay time of methane/air calculated using detailed chemical models and measured experimentally is much longer than that for simplified chemical models. This means that for the same initial conditions a successful detonation initiation through the temperature gradient requires much shallower gradient, and the critical size $L_{cr}$ of the hot pocket should be much larger. As a representative example, Figure 12 shows evolution of the temperature and pressure profiles calculated using DRM19 chemical model for same initial conditions as for simplified models in Figures 6, 9. A steady detonation was not observed in simulations for shallower gradients up to $L < 30\,\text{cm}$.

In the case of detailed chemistry the initiating reactions proceed without heat release, and the gas-dynamic perturbations at the induction stage are very weak. The wave of exothermal reaction follows the spontaneous wave path with the delay determined by the time scale of termination reactions. Therefore, even for sufficiently shallow temperature gradient, so that the spontaneous wave initially propagates supersonically, for detailed chemical model DRM19 there is no detonation, while simplified chemical models can yield successful detonation even for much steeper gradients. The temporal evolution of temperature and pressure profiles in Fig.12 shows the formation of a combustion wave behind the weak shocks, which run ahead of the reaction wave front. As it is seen in Fig. 12, the velocity of the spontaneous reaction wave at the minimum point, where the pressure wave overtakes the reaction wave, is not sufficient to sustain synchronous amplification of the pressure pulse in the flow behind the shock. The pressure wave runs ahead of the reaction wave and the velocity of the reaction wave decreases.

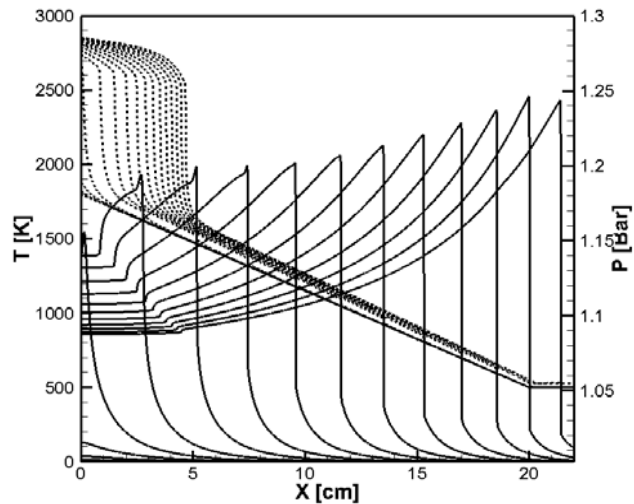

**Figure 12:** Evolution of the temperature (dashed lines) and pressure (solid lines) profiles calculated for DRM19 model at $P_0 = 1 bar$ and the same conditions as in figures 6, 9.

At high pressures the length of the induction stage becomes about the length of the chain termination stage at temperature $T \sim 1200K$, and the upper part of the gradient, where detonation can develop behind the shock wave, decreases. Therefore, one would expect that the implementation of steady detonation for the same temperature gradient can occur at a higher initial pressure. Figures 13 shows that for the same temperature conditions as in Figs. 6 and 9, but for initial pressure $P_0 = 5 bar$, the pressure waves overtaking the reaction wave are not sufficient to sustain synchronous amplification of the pressure pulse in the flow behind the shock. The pressure waves run ahead of the reaction wave and the developing detonation quenches.

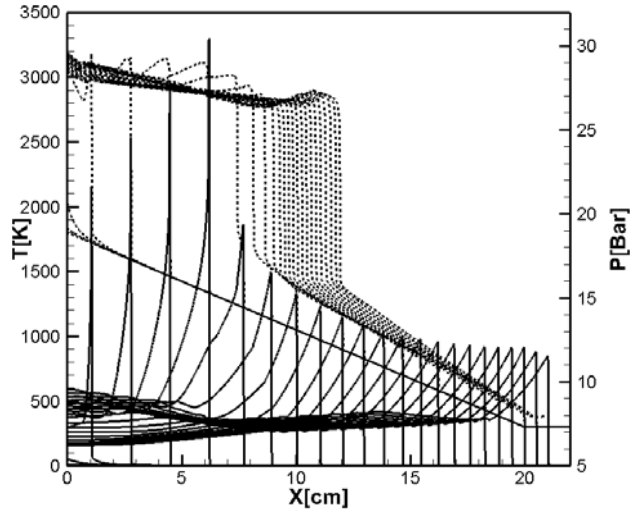

**Figure 13:** Evolution of the temperature (dashed lines) and pressure (solid lines) calculated for DRM19 model at $P_0 = 5\text{bar}$ and for the same $T^* = 1800\text{K}$, $T_0 = 300\text{K}$ as in Figs. 6, 9.

Of interest is the initiation of a combustion wave by a temperature gradient at high ambient temperatures outside the gradient. Radulescu et al. [51] noted that as more uniform reaction zone, as stronger reaction is coupled with the shock wave. Figure 14 shows the time evolution of the temperature and pressure profiles calculated using DRM19 model for initial pressure $P_0 = 10\text{bar}$ and temperature outside the hot spot, $T_0 = 500\text{K}$.

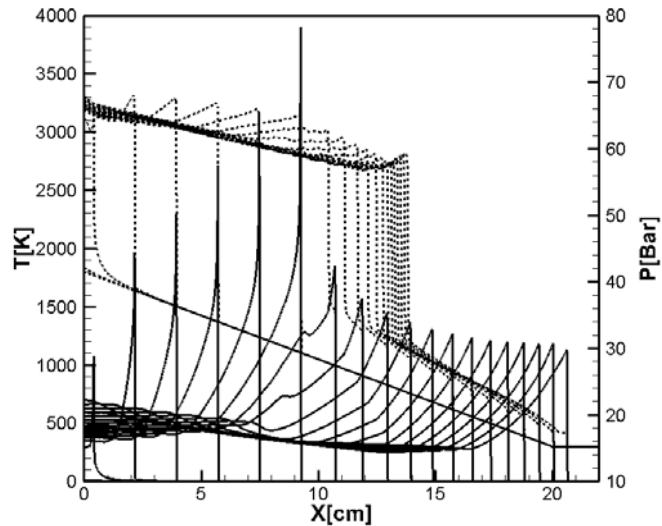

**Figure 14:** Evolution of the temperature (dashed lines) and pressure (solid lines) calculated for DRM19 model at $P_0 = 10\text{bar}$, $T^* = 1800\text{K}$, $T_0 = 300\text{K}$.

In this case, even when the thermal runaway propagates supersonically the thermal runaway path differs considerably from the spontaneous wave path. From Figs. 12 and 13 one can observe that the pressure waves overtake the reaction wave, run away and do not provide synchronous amplification of the pressure pulse and reaction wave in the flow behind the shock even for a very shallow temperature gradient $L > 20$cm cm and for pressure $P_0 = 5$bar.

Although the expected sizes of the temperature gradient shown in Fig. 15 according to Eq.(15) decrease with increasing initial pressure, it was found that the developing detonation quenches (Fig. 14) for all values of $T_0 < 1100$K in spite of rather large ($L = 20$cm) size of the hot pocket.

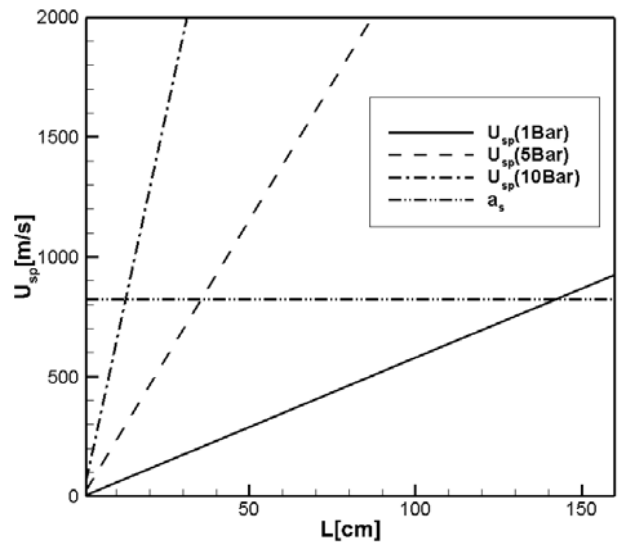

**Figure 15:** The points of intersection with the sound speed show the minimum lengths $L_{cr}$ of the temperature gradients at initial pressure 1atm, 5atm, 10atm (the steepest gradients $L = L_{cr}$, initiating detonation for $T_0 = 300$K.

For a sufficiently high ambient temperature $T_0$ there is no hydrodynamic resistance outside the gradient since outside the gradient the wave propagates at constant ambient density, and therefore the transition to detonation occurs for a steeper gradient (smaller $L_{cr}$) as it is shown in Fig. 15, where a steady CJ-detonation is developed for a steeper gradient with the critical size of the hot pocket $L_{cr} = 6$cm and $T_0 = 1100$K.

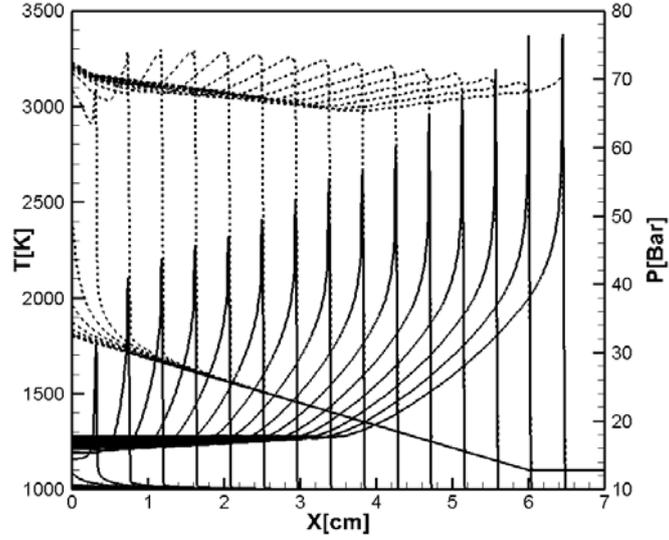

**Figure 16:** Evolution of the temperature (dashed lines) and pressure (solid lines) for developing a steady detonation calculated for DRM19 model at $P_0 = 10\,\text{bar}$, $T_0 = 1100\,\text{K}$.

In a sense, the high temperature outside the gradient region is equivalent, but not completely, to a shallower gradient. The induction stage, which is distinctive for chain reactions can be "skipped" at sufficiently high ambient temperatures, and the scenario of a detonation wave formation at the end of the gradient or outside the gradient becomes more complicated. At very high pressures, about 40-50atm, and for steeper gradients the intensity of the shock wave can be not sufficient for a local explosion to occur during the induction time in the mixture behind the shock. In this case the volume thermal explosion ahead of the shock wave can occur after or before the developing detonation leaves the temperature gradient entering the region of ambient temperature. Depending on the relation between the time of detonation formation and the time of the thermal explosion development, the detonation wave can "meet" with a thermal explosion at the end or outside the temperature gradient.

To analyze the difference between the detailed and one-step models we consider the thermal sensitivity of the induction time $\beta = -(T/\tau_{ind})(\partial \tau_{ind} / \partial T)$. Approximating the induction time as

$$\tau_{ind} = A\exp(E_a / RT), \tag{16}$$

and taking into account that $(\partial \tau_{ind} / \partial T) = -\tau_{ind}(E_a / RT^2)$, we obtain $\beta = (E_a / RT)$, where $E_a$ can be considered as an effective global activation energy, which is different for a one-step and for detailed chemical models. Figure 117 shows $\beta(T)$ calculated for the one-step model (Eq. (2)) and for DRM19 model for different pressure: $P_0 = 1, 5,$ and $10\,\text{bar}$. It can be seen that with increasing initial pressure, $\beta_{DRM}$ decreases and approaches $\beta_{1-step}$ for a one-step model.

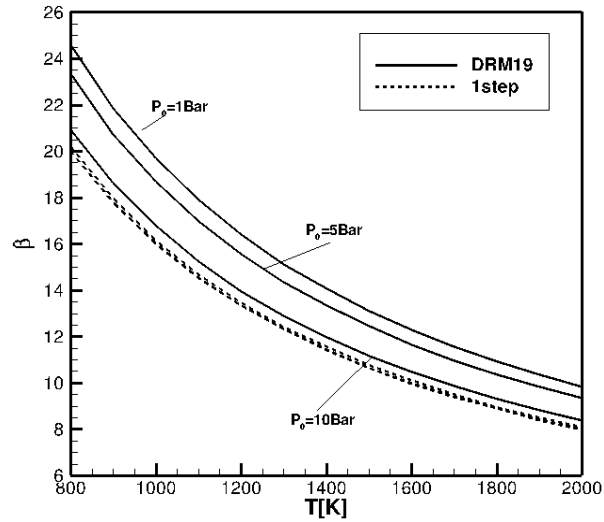

**Figure 17:** $\beta(T)$ calculated for the one-step model (dashed lines) and for DRM19 model (solid lines) for different pressure: $P_0 = 1\text{bar}$, $P_0 = 5\text{bar}$, $P_0 = 10\text{bar}$.

From this behavior of the thermal susceptibility, it can be assumed that the detonation initiation scenario for the detailed model at high pressures will be somewhat more similar to that realized for a one-step model. Since at high pressure the induction time is considerably smaller, the minimal steepness of the gradients necessary for the implementation all combustion modes, in particular, for detonation is considerably increased ($L_{cr}$ decreases). However, since both values $\tau_{ind}$ and $(\partial \tau_{ind} / \partial T)$ remain for the detailed DRM19 model at least two orders of magnitude smaller than for a one-step model, the conditions for the detonation initiation for detailed chemistry also remain considerably different and complicated.

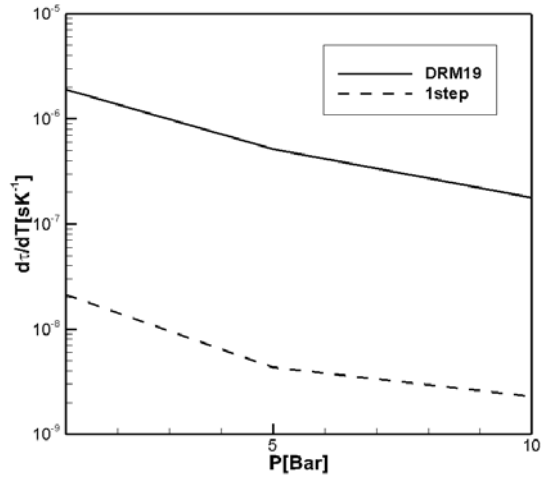

**Figure 18:** $(\partial \tau_{ind} / \partial T)$ vs pressure calculated for the one-step model (dashed lines) and for DRM19 model (solid lines).

## 4. Conclusions

In the present study we investigated the influence of different chemical models on the initiation detonation by the initial temperature gradient. We examined various chemical models, simplified chemical schemes: 1) a one-step Arrhenius model, 2-step and 4-step (J-L) models, and compared these models with the detailed multi-step 35 reactions scheme and DRM-19 model. All the models were calibrated to correctly reproduce the key features of laminar flames, including velocity, the width of the flame front and adiabatic temperature. The ignition delay times and the pressure dependence of the laminar flame velocities have been studied for all chemical models and validated against standard GRI 3.0-mech and experimental data for a wide range of temperatures and pressures. Simulations of the standard hot spot problem with linear temperature gradient were performed using high-resolution numerical simulations to investigate the scales of the initial temperature nonuniformity required for initiation of different combustion modes with special attention given to conditions required to development a detonation. The necessary condition for initiation detonation by the temperature gradient, which gives a minimum size of the hot pocket, was obtained and found to be in good agreement with numerical simulations. It was shown that the requirements in terms of the temperature gradient steepness (size of the hot spot) and of the initial temperature outside the hot pocket to trigger detonation are considerably different for the simplified and for the detailed chemical models. For the detailed chemical model, the size of the temperature gradient and corresponding size of the hot spot that could initiate detonation via the gradient mechanism are in orders of magnitude greater than that for simplified models. As the pressure increases, the required for detonation size of the hot spot decreases. However, for a detailed chemical model, the steady detonation can be formed only at high temperatures outside the hot spot, such that there can be competition between the detonation developing on the gradient and the thermal explosion, which develops at the gradient end or outside the hot spot.

Using a one-step model for simulations of flame dynamics and DDT one has to make a choice of how to calibrate the model parameters, which depends on the particular application. For a stationary planar flame, one can calibrate the model parameters to fit the normal flame velocity and the flame thickness, obtaining at least a qualitatively correct description of the flame dynamics. The situation is different for nonstationary processes, like ignition, transient processes, and the like. In a trivial way, taking e.g., an unrealistically low activation energy in order to match more or less accurately the induction times obtained from the detailed chemical model or measured experimentally, the parameters of the simplified (one-step or two-step) models can be chosen in order to reproduce the constant-volume induction times. However, this choice leads to incorrect value of activation energy, which is several times smaller than the global activation energy.

One of the conclusions of the present studies is that the rather large (6-7cm) size of the hot spot and the high temperature outside the hot spot required to initiate detonation by the spontaneous reaction wave in the gradient mechanism mean that when simplified chemical models are used in DDT simulation the results of simplified models must be interpreted with special caution. To initiate a detonation, the temperature gradient must be shallow enough to build a strong pressure peak, which is of the magnitude of the von Neumann pressure peak in a detonation wave. It was shown that for high pressures the gradient steepness for direct initiation of detonation decreases considerably, which means that at very high pressure the detonation can be ignited by a smaller initial nonuniformity, which is of substantial practical interest for risk assessment to minimize accidental explosions, in particular, for safety guidelines in industry and for better understanding knock in a very high compression SI-engines. The results of this work can be a good platform for further multidimensional study of DDT and for experimental techniques to produce a detonation by a temperature gradient without strong initial hydrodynamic perturbations.

**Acknowledgments**

This work was supported by the National Natural Science Foundation of China under grants 11732003 and 11521062, by National Key R&D Program of China (No. 2017YFC0804700), and by funds of the opening project number KFJJ17-08M of State Key Laboratory of Explosion Science and Technology, Beijing Institute of Technology.

The authors are gratefully acknowledge insightful comments made by P. Clavin during the review process. One of the authors (Misha Liberman) acknowledges helpful discussions with many friends and colleagues. Especially I would like to thank Mike Kuznetsov, Grisha Sivashinsky and Igor Rogachevskii for their collaborations and deep insights into problems presented in this paper. Special thanks always to Dr. Koshatsky (Inna) for the wonderful environment, unfailing companionship and valuable advice.

**Disclosure statement**

No potential conflict of interest was reported by the authors.

**Appendix A**

In direct simulation of multi-species reactive flows, the flow and chemical time scales may be comparable, but the spatial resolution required for chemistry demands fine grids in order to resolve the thin reaction zones. The time step allowed when integrating the governing equations with explicit schemes is controlled by the diffusive stability limit dictated by these few chemical species. Resolution and convergence tests were thoroughly performed to ensure that the resolution is adequate to describe and to capture details of the problem in question and to avoid computational artifacts. This is especially important for a mixture consisting of many species with a large number of reactions in the case of a detailed chemical model Figures (1, 2, 3, 4) show the resolution and convergence tests for the structure and velocities of a planar laminar flames in simulations with simplified and detailed chemical schemes. The convergence of the solution is quite satisfactory already for 8 grid points per flame width at initial pressure 1atm. The flame velocity, density, and temperature for this resolution differ by less than 2-3% from the converged solution, while a resolution of 16 computational cells results in an "exact" solution. A grid spacing independence was verified up to the resolution of 32 grid points per width of the flame and higher, which corresponds to the cell size less than 10-12μm. 2D simulations of a planar flame in the present work predicted the same values of burning velocity as 1D simulations.

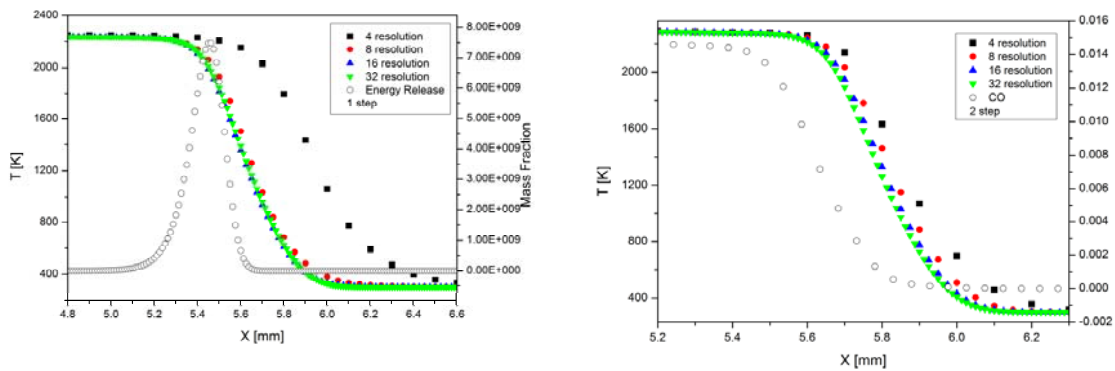

**Figure A1(a, b):** Resolution tests for 1-step and 2-step models at normal ambient conditions ($T_0 = 300$ K, $P_0 = 1$atm).

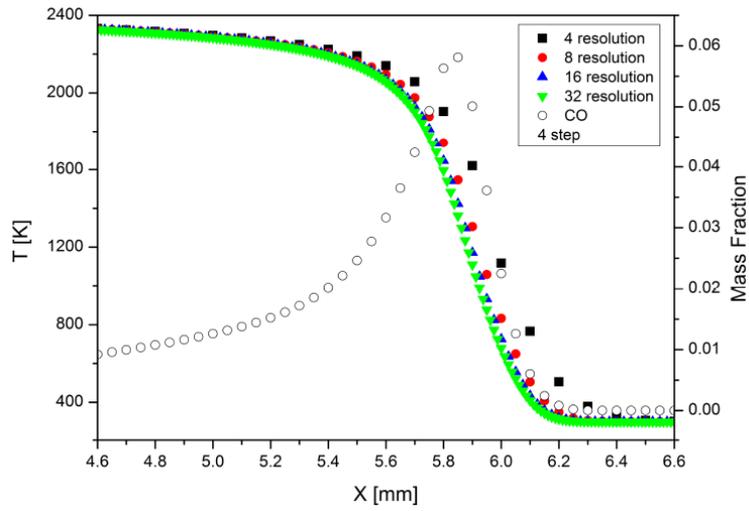

**Figure A2:** Resolution tests for 4-step model at normal ambient conditions ($T_0 = 300$ K, $P_0 = 1$ atm).

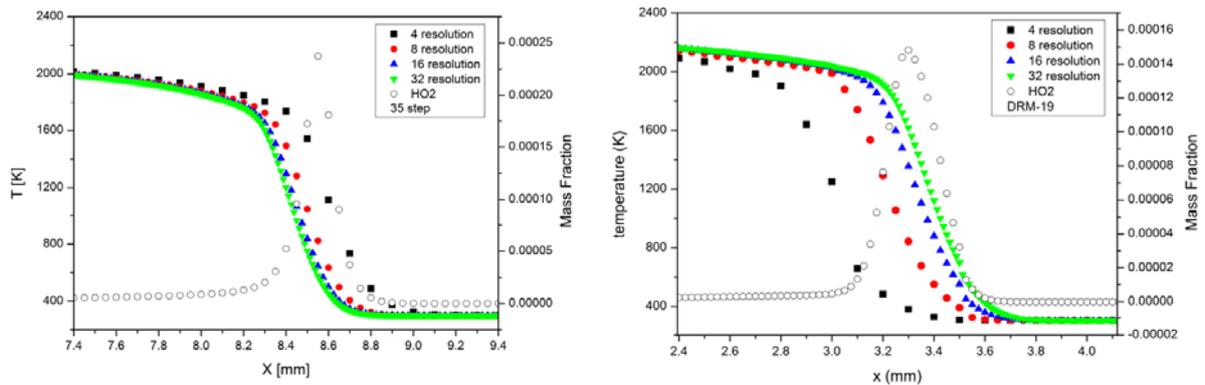

**Figure A3 (a, b):** Resolution tests for Smooke 35-step [25] and DRM-19 (84 reactions) models at normal ambient conditions ($T_0 = 300$ K, $P_0 = 1$ atm).

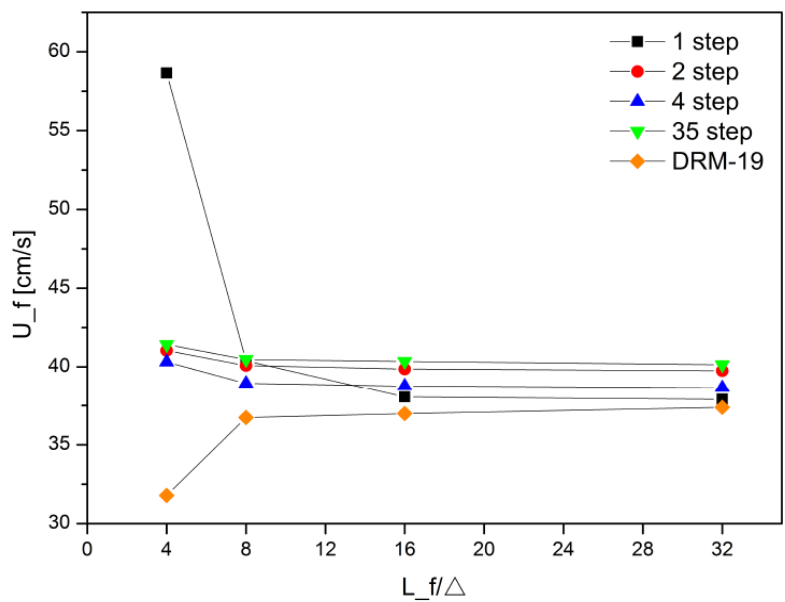

**Figure A4:** Convergence of solutions for simplified and detailed chemical models.

The most demanding is the case where the detonation arises as a result of auto-ignition inside the hot-spot. Therefore one should appropriately resolve the combustion waves propagating through reacting medium on the background of elevated temperature and pressure behind the shock front. Besides the coupling of the reaction wave and shock should be resolved taking into account that the flame thickness is much larger than the width of the shock front. According to this verification of the solver was performed using a common test for the accuracy of computational fluid codes, which was heavily investigated by Sod [48], in which the result of the numerical solution of one-dimensional Riemann problem is compared with the analytical solution. Figures A5(a, b) show solution for the Sod problem (Fig. 5a) and numerical solution to the problem (Fig. A5b).

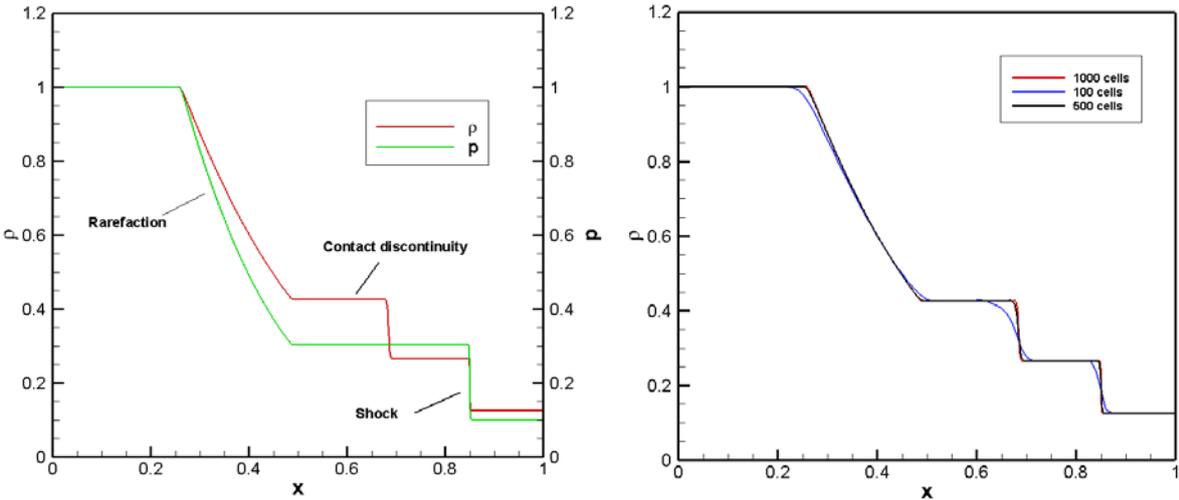

**Figure A5 (a, b):** Analytical solution (5a) for the Sod problem and numerical solution to the problem (Fig. 5b).

It is seen from Fig. A5b that the converged solution for the detonation initiation problem requires to use a finer resolution from the very beginning. At elevated pressures the flame thickness decreases. On the other hand the diffusivity of the numerical scheme smoothens the shock front over 5 computational cells. Therefore no artificial coupling is possible for the chosen meshes determining fine resolution.